\documentclass[a4paper,11pt]{nveart}
\usepackage{graphicx,amsmath,amssymb}

\begin{document}

\company{Accepted for publication in}
\journal{Astroparticle Physics}

\begin{frontmatter}
\title{Implementation of a Gauss convoluted Pandel PDF\\
       for track reconstruction in Neutrino Telescopes}

\author[Utrecht]{N. van Eijndhoven},
\author[Atlanta]{O. Fadiran} and
\author[Atlanta]{G. Japaridze}

\address[Utrecht]{Department of Physics and Astronomy,
                  Utrecht University, Utrecht, The Netherlands}
\address[Atlanta]{Center for Theoretical Studies of Physical Systems,
                  Clark Atlanta University, Atlanta, U.S.A}

\begin{abstract}
A probability distribution function is presented which provides a
realistic description of the detection of scattered photons.
The resulting probabilities can be described analytically
by means of a superposition of several special functions.
These exact expressions can be evaluated numerically
only for small distances and limited time residuals, due
to computer accuracy limitations.
In this report we provide approximations for the exact
expressions in different regions of the distance-time
residual space, defined by the detector geometry and
the space-time scale of an event.
These approximations can be evaluated numerically with
a relative error with respect to the exact expression
at the boundaries of less than $10^{-3}$.
\end{abstract}

\end{frontmatter}

\section{Introduction}
In the track reconstruction process in neutrino telescopes it is
customary to perform a log-likelihood minimization
using a gamma function based Probability Distribution Function (PDF).
The use of a so-called Pandel PDF \cite{pand} has been investigated
in the reconstruction of muon tracks in the AMANDA detector \cite{mureco}.
However, the problem with a Pandel PDF is that it cannot cope with negative
time residuals which may result due to time jitter in the detection devices.
One way to overcome this limitation is to consider a convolution of a Pandel
PDF with a Gaussian, the latter accounting for the finite time resolution of the detector.
The resulting expression, referred throughout this report as a CPandel PDF,
can be evaluated analytically.
This involves the superposition of various
special functions, which are available in the GNU Scientific
Library (GSL) \cite{gnu}.
However, due to computer accuracy limitations, the numerical evaluation
of a CPandel PDF is restricted to a rather limited distance-time residual
domain.
In this report we present analytical approximations which are necessary to 
extend the applicability domain of a CPandel PDF for practical cases.
The implementation shown here is tailored for the geometry of the
AMANDA neutrino telescope \cite{mureco} and has been realized
in an analysis framework (IcePack) based on ROOT \cite{root}
and an analysis toolbox (Ralice) \cite{ralice} which was
originally developed for the Alice experiment at the future CERN-LHC collider.

\section{The CPandel probability distribution function}
In a reconstruction procedure, track parameters emerge as the solution of an optimization problem:
given experimentally measured values (in our case photon detection times $t_{hit}$),
find the values of parameters minimizing the log of the likelihood $\mathcal{L}$.
The latter is given by \cite{mureco}
\begin{equation}
\label{eq:ll}
\mathcal{L}=\prod_{j}\,p(a,\,t_{hit,j}) ~,
\end{equation}
where $a$ denotes the parameter(s) characterizing the hypothesis of an event and
the PDF $p(a,t_{hit,j})$ describes a photon arrival at the receiver $j$.
It is convenient to use the time residual $t$ as a PDF variable
\begin{equation}
\label{eq:t}
t \equiv t_{hit}-t_{geom} ~,
\end{equation}
where $t_{geom}$ is the photon arrival time for a case of no scattering and no absorption.\\
Note that in our notation we have dropped the index $j$, since for the scope of the present report
it is sufficient to limit ourselves to a one source-one receiver situation.

\begin{figure}[htb]
\begin{center}
\includegraphics[keepaspectratio,width=10cm]{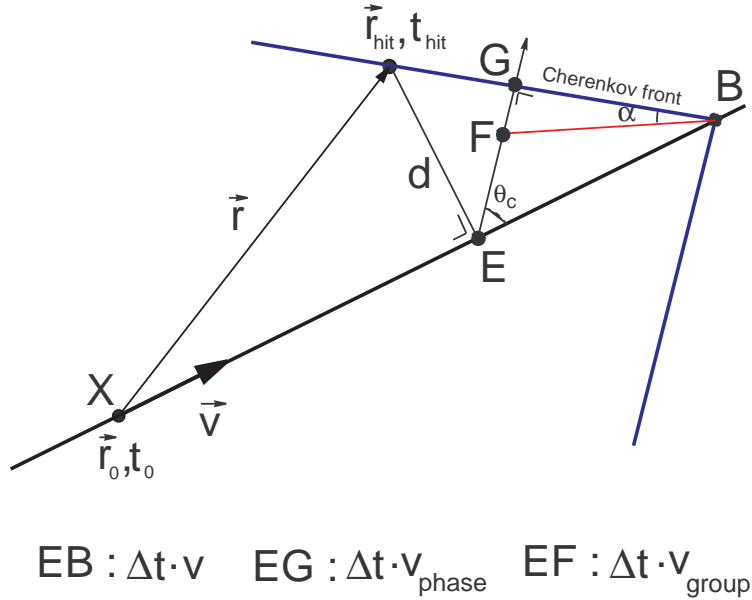}
\end{center}
\caption{Geometry of the signal generation process.
         Further details can be found in the text.}
\label{fig:tgeom}
\end{figure}

The generation of detector signals corresponds to a process in which a charged particle moves
along a straight track with velocity $v$ exceeding the in-medium light speed $c_{m}$.
This gives rise to emission of Cherenkov radiation, which is recorded by the detection devices.
The geometrical situation is sketched in Fig.~\ref{fig:tgeom}, in which the particle starts
from a point $X$ at $(\vec{r}_{0},t_{0})$ and arrives at a point $B$ when the Cherenkov front
hits the detector at $(\vec{r}_{hit},t_{hit})$.
The track point labeled $E$ indicates the point of closest approach w.r.t. the receiver location.
The geometrical (expected) arrival time can be evaluated by considering
Cherenkov photons emitted from the particle track and propagating freely
(without scattering and absorption) to arrive at the receiver at $(\vec{r}_{hit},t_{hit})$.
This specific case implies that $t_{geom}=t_{hit}$.

Indicating by $\Delta t$ the time it takes for the particle to travel from $E$ to $B$,
it is seen from Fig.~\ref{fig:tgeom} that in the same time interval the Cherenkov front
has moved with phase velocity $v_{phase} \equiv c/n_{ph}$ from $E$ to $G$.
Here $c$ indicates the light speed in vacuum and $n_{ph}$ is called the phase refractive index.
However, a detector is sensitive for real photons which travel with the group
velocity $v_{group} \equiv c/n_{gr}$, where $n_{gr}$ is called the group refractive index.\\
Since $v_{group}<v_{phase}$, the real photon signal lags behind the Cherenkov front.
In Fig.~\ref{fig:tgeom} this is indicated by the line element $EF$, traveled by the real Cherenkov
photons in the time interval $\Delta t$ mentioned above.
So, one can regard the real photons to be located on a wavefront given by $BF$
which is comparable to the Cherenkov front provided the complement of the Cherenkov
angle $\theta_{c}$ is reduced by $\alpha$.\\
Assuming $v=c$ we have $\cos(\theta_{c})=1/n_{ph}$. The expected arrival time is then given by 
\begin{equation}
\label{eq:tgeom}
t_{geom}=t_{0}
    +\frac{1}{c}\biggl [\hat{v}\cdot\vec{r}+d\,\frac{n_{gr}n_{ph}-1}{\sqrt{n^{2}_{ph}-1}}\biggr] ~,
\end{equation}
where $\hat{v}$ indicates the unit vector in the moving direction of the particle.

Minimization w.r.t. the time residuals $t=t_{hit}-t_{geom}$ for all observed signals
provides the basis for track reconstruction, as mentioned before.\\
To achieve this, a Pandel PDF $p(\xi,\,\rho,\,t)$ was suggested in \cite{pand}~:
\begin{equation}
\label{pandel}
p(\rho,\xi,t)=\frac{\rho^{\xi}t^{\xi-1}}{\Gamma(\xi)}\,e^{-\rho t} ~,
\end{equation}
where $\xi$ and $\rho$ are phenomenological parameters related to the characteristics of the medium.\\
The parameter $\xi$ represents the distance between the emission and detection locations
of a Cherenkov photon in units of the mean photon scattering length $\lambda$,
i.e. $\xi=d/(\lambda\sin\theta_{c})$.
Based on the experiences from the  muon track reconstruction procedure with the
AMANDA Neutrino Telescope \cite{mureco}, we use the following parameter values
throughout this report
\begin{equation}
\label{parms}
\lambda\,=\,33.3\,\mathrm{m},\quad\rho=0.004\,\mathrm{ns}^{-1} ~.
\end{equation}

A PDF for realistic signals should account for the finite time resolution of the detector.
This can be achieved by convolving a Pandel PDF $p$ (which describes only the propagation
of photons in a medium, i.e. an ideal measurement situation of a detector with a zero time resolution)
with a time jitter function.
The latter may be described by a gaussian with a mean of zero and standard deviation $\sigma$.
The width of the gaussian distribution, $\sigma$, represents the time resolution of the detector
and usually is made up from various sources, which motivates the use of a gaussian.
We call this PDF CPandel and denote it as $\mathcal{F}_{\sigma}$~:
\begin{equation}
\label{0}
\mathcal{F}_{\sigma}(\rho,\xi,t)=
\int^{\infty}_{0}\,\frac{{\rm d}x}{\sqrt{2\pi\sigma^{2}}}\,p(\rho,\,\xi,\,x)\,e^{-(t-x)^{2}/2\sigma^{2}} ~.
\end{equation}

The convoluted PDF $\mathcal{F}_{\sigma}(\rho,\xi,t)$ can be calculated exactly \cite{GM}~:
\begin{equation}
\label{cpexact}
\mathcal{F}_{\sigma}(\rho,\,\xi,\,t)=
\frac{\rho^{\xi}\sigma^{\xi-1}e^{-t^{2}/2\sigma^{2}}}{2^{(1+\xi)/2}}
\,\Biggl[ \frac{_{1}F_{1}(\frac{1}{2}\xi,\frac{1}{2},\frac{1}{2}\eta^{2})}{\Gamma(\frac{1}{2}(\xi+1))}
-\sqrt{2}\,\eta\;\frac{_{1}F_{1}(\frac{1}{2}(\xi+1),\frac{3}{2},\frac{1}{2}\eta^{2})}{\Gamma(\frac{1}{2}\xi)}\Biggr] ~,
\end{equation}
where
\begin{equation}
\label{eta}
\eta=\rho\sigma-\frac{t}{\sigma}
\end{equation}
and $_{1}F_{1}$ is the confluent hypergeometric function \cite{AS}.
The latter is implemented in the GNU Scientific Library (GSL) \cite{gnu}, which enables
numerical evaluation of $\mathcal{F}_{\sigma}(\rho,\xi,t)$.

\newpage

\begin{figure}[htb]
\begin{center}
\includegraphics[keepaspectratio,width=11cm]{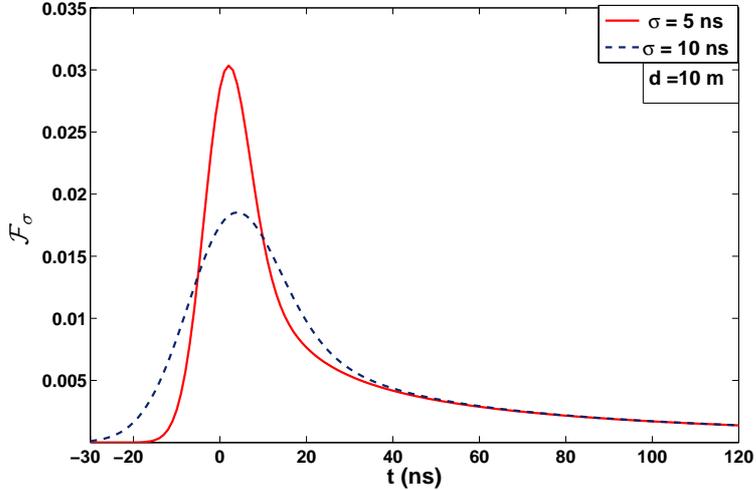}
\end{center}
\caption{CPandel PDF for two values of the time resolution parameter $\sigma$ ($5$ and $10$ ns).
         The distance of closest approach $d$ is 10 m.}
\label{fig:cpandel}
\end{figure}

Accounting for the detector finite time resolution provides a solution of the problem
of negative time residuals~: $\mathcal{F}_{\sigma}$ exists for any $t$ and causality
is satisfied. For $t<0$ the finite value of the CPandel PDF is purely due to the detector
time jitter \cite{GM}.
A typical profile of $\mathcal{F}_{\sigma}$ is shown in Fig.~\ref{fig:cpandel}.

In a reconstruction procedure the minimizer moves randomly within the $\xi\,-\,t$
(i.e. distance-time residual) plane.
As such, it is essential to maximize the area in which numerical evaluations of
$\mathcal{F}_{\sigma}$ can be performed.\\
Although $\mathcal{F}_{\sigma}$ possesses all the desirable attributes for a PDF \cite{GM}
and $_{1}F_{1}$ is implemented in GSL, computer accuracy limitations pose a problem in minimizing
the log-likelihood based on $\mathcal{F}_{\sigma}$.
For instance, the attempt to calculate $\mathcal{F}_{5}(0.004,2,190)$ fails due to overflow
in the GSL procedure.
Consequently, the GSL implementation of $_{1}F_{1}$ does not allow evaluation of
$\mathcal{F}_{\sigma}$ when the distance is 66.6 m and the time residual is 190 ns.
As such, evaluation of the exact expression of $\mathcal{F}_{\sigma}$ as given by \eqref{cpexact}
cannot be performed in all the areas relevant for realistic cases.\\
Due to these limitations, it is necessary to find an approximation which
allows numerical evaluation of the CPandel PDF in all physically relevant areas,
while retaining all the qualities defined from the exact expression \eqref{cpexact}.
Our strategy is to use \eqref{cpexact} in the area where the GSL implementation of
$_{1}F_{1}$ yields reliable results and outside that area use an approximation which
allows extension of the support for the minimizer procedure.

Since the analytic properties of $_{1}F_{1}$ are well established \cite{AS},
it is straightforward to find approximations for $\mathcal{F}_{\sigma}$ in terms of uniform
(i.e. numerically convergent) expansions
\footnote{Note that using an asymptotic expansion (with regard to a variable, say $t$)
for the CPandel PDF may disturb the approximation when another variable ($\xi$) is varied \cite{GM}.}.\\
Depending on the position in the $\xi-t$ plane (as the minimizer moves away from the origin),
different functions have to be used to approximate $\mathcal{F}_{\sigma}$.
However, it should be noted that these different expressions for the CPandel PDF
(corresponding to different regions in the distance-time plane) represent the same function
approximated by means of analytic continuation and uniform expansion.
In other words, the results presented hereafter are not originating from
introducing any new ad hoc functions or parameters.\\
Below we present the various expressions for the CPandel PDF corresponding to different relevant regions
of the distance-time residual plane.

\section{Expressions for $\mathcal{F}_{\sigma}$ in different distance-time regions}
For direct hits (i.e. small $\xi$ and small $|t|$), we can use the exact expression for
$\mathcal{F}_{\sigma}$ as given by eq.~\eqref{cpexact}, since this is correctly handled
by the GSL implementation on this domain.\\
To extend the coverage to all relevant areas in the $\xi-t$ plane, it is necessary and sufficient
to consider the following regimes for the variables $\xi$ and $\eta$ (for $\eta$ see \eqref{eta})~:
positive and negative $\eta$ (corresponding to $t<\rho\sigma^{2}$ and $t>\rho\sigma^{2}$, respectively),
$\xi \leq 1$ and $\xi>1$.
Approximations for $\mathcal{F}_{\sigma}$ in the latter four regions exhaust all the possibilities
following from the analytic continuation procedure of the function $_{1}F_{1}$, of which
the details can be found at the end of this report.\\
Consequently, the support for $\mathcal{F}_{\sigma}$ consists of the following five parts~:\\
\begin{itemize}
\item[] Region 1 : direct hits region - small $\xi$ and small $|t|$.
\item[] Region 2 : large positive $t$ and restricted $\xi$.
\item[] Region 3 : large $\xi$ and $\eta \leq 0$ (i.e. $t \geq \rho\sigma^{2}$).
\item[] Region 4 : large $\xi$ and $\eta \geq 0$ (i.e. $t \leq \rho\sigma^{2}$).
\item[] Region 5 : large negative $t$ and restricted $\xi$.\\
\end{itemize}
In the following we denote the CPandel PDF in region $j$ by $f_{j}$.

\subsection*{Zero distance}
On the $t$-axis, i.e. where $\xi=0$, the exact expression \eqref{cpexact} turns into
\begin{equation}
\label{ksizeri}
\mathcal{F}_{\sigma}(\rho,0,t)=\frac{e^{-t^{2}/2\sigma^{2}}}{\sqrt{2\pi\sigma^{2}}} ~,
\end{equation}
which we use for $\xi=0$.

\subsection*{Region $1$~: small $\xi$, small $|t|$}
Here GSL allows the use of the exact expression \eqref{cpexact}.
In other words, in the region corresponding to direct hits no approximation is used.\\
In this region, for $t \leq 0$ we use the constraint
\begin{equation}
\label{tneg1}
-5\sigma \leq t \leq 0 ~,
\end{equation}
which yields satisfactory results in matching the values of $\mathcal{F}_{\sigma}$
calculated via the exact expression \eqref{cpexact} with the values of $f_{4}$ and $f_{5}$
(as outlined below). 

\subsection*{Region $2$~: $\xi \leq 1$ and large positive $t$ (i.e. $t \gg \sigma$)}
In this region, $\mathcal{F}_{\sigma}$ is approximated by
\begin{eqnarray}
\label{101}
\mathcal{F}_{\sigma}(\rho,\xi,t)=e^{\rho^{2}\sigma^{2}/2}\,p(\rho,\xi,t) \equiv f_{2}(\rho,\xi,t) ~.
\end{eqnarray}
Here $p(\rho,\xi,t)$ is the Pandel PDF as given in eq.~\eqref{pandel}.

\subsection*{Region $3$~: $\xi \geq 1$ and $t \geq \rho\sigma^{2}$}
In this region $\mathcal{F}_{\sigma}$ is approximated by
\begin{equation}
\label{rightdef}
\mathcal{F}_{\sigma}(\rho,\xi,t)=\frac{e^{\alpha}}{\Gamma(\xi)}\,\Phi \equiv f_{3}(\rho,\xi,t) ~,
\end{equation}
where
\begin{eqnarray}
\nonumber
\alpha &= & -\frac{t^{2}}{2\sigma^{2}}\,+\,\frac{\eta^{2}}{4}\,-\,\frac{\xi}{2}\,+\,\frac{1}{4}\,+\,k(2\xi-1)\,-\,\frac{1}{4}\,\ln(1+z^{2})\,-\,\frac{\xi}{2}\,\ln(2)\cr\cr
& + &\frac{\xi-1}{2}\,\ln(2\xi-1)\,+\,\xi\ln(\rho)\,+\,(\xi-1)\,\ln(\sigma) ~,
\end{eqnarray}
\[
k\,=\,\frac{1}{2}\,\Biggl[z\,\sqrt{1\,+\,z^{2}}\,+\,\ln\biggl(z\,+\,\sqrt{1\,+\,z^{2}}\biggr)\Biggr],
\quad z\,=\,-\,\frac{\eta}{\sqrt{4\xi-2}}\,>\,0 ~,
\]
\vspace{2mm}
\[
\Phi\,=\,1\,-\,\frac{N_{1}}{(2\xi-1)}+\,\frac{N_{2}}{(2\xi-1)^{2}} ~,
\]
\[
N_{1}\,=\,\frac{\beta}{12}\,(20\beta^{2}\,+\,30\beta\,+\,9),
\quad N_{2}\,=\,\frac{\beta^{2}}{288}\,(6160\beta^{4}\,+\,18480\beta^{3}\,+\,19404\beta^{2}\,+\,8028\beta\,+\,945) ~,
\]
\[
\beta\,=\,\frac{1}{2}\,\biggl(\frac{z}{\sqrt{1+z^{2}}}\,-\,1\biggr) ~.
\]

\subsection*{Region $4$~: $\xi \geq 1$ and $t \leq \rho\sigma^{2}$}
In this region, $\mathcal{F}_{\sigma}$ is approximated by
\begin{equation}
\label{leftdef}
\mathcal{F}_{\sigma}(\rho,\xi,t)=
\frac{\rho^{\xi}\sigma^{\xi-1}\,e^{-t^{2}/2\sigma^{2}+\eta^{2}/4}}{\sqrt{2\pi}}
\,U(\xi)\,e^{-k\,(2\xi-1)}\,(1+z^{2})^{-1/4}\,\Psi \equiv f_{4}(\rho,\xi,t) ~,
\end{equation}
where
\[
U(\xi)\,=\,e^{\xi/2-1/4}\,(2\xi-1)^{-\xi/2}\,2^{(\xi-1)/2} ~,
\]
\vspace{2mm}
\[
k\,=\,\frac{1}{2}\,\Biggl[z\,\sqrt{1\,+\,z^{2}}\,+\,\ln\biggl(z\,+\,\sqrt{1\,+\,z^{2}}\biggr)\Biggr] ~,
\]
\vspace{2mm}
\[
\Psi\,=\,1\,+\,\frac{N_{1}}{(2\xi-1)}+\,\frac{N_{2}}{(2\xi-1)^{2}} ~.
\]
$N_{j}$, and $\beta$ are the same as for region 3 and $z$ is now defined as
\[
z\,=\,\frac{\eta}{\sqrt{4\xi-2}}\,>\,0 ~.
\]

\subsection*{Region $5$~: $\xi \leq 1$ and $t \ll -\sigma$}
In this region $\mathcal{F}_{\sigma}$ is approximated by 
\begin{equation}
\label{ll}
\mathcal{F}_{\sigma}(\rho,\xi,t)=
\frac{(\rho\sigma)^{\xi}}{\sqrt{2\pi\sigma^{2}}}\,\eta^{-\xi}\,e^{-t^{2}/2\sigma^{2}}
 \equiv f_{5}(\rho,\xi,t) ~.
\end{equation}

\section{Description of the support of $\mathcal{F}_{\sigma}$ in the $\xi-t$ plane}
\begin{figure}[htb] 
\begin{center}
\includegraphics[keepaspectratio,width=12cm]{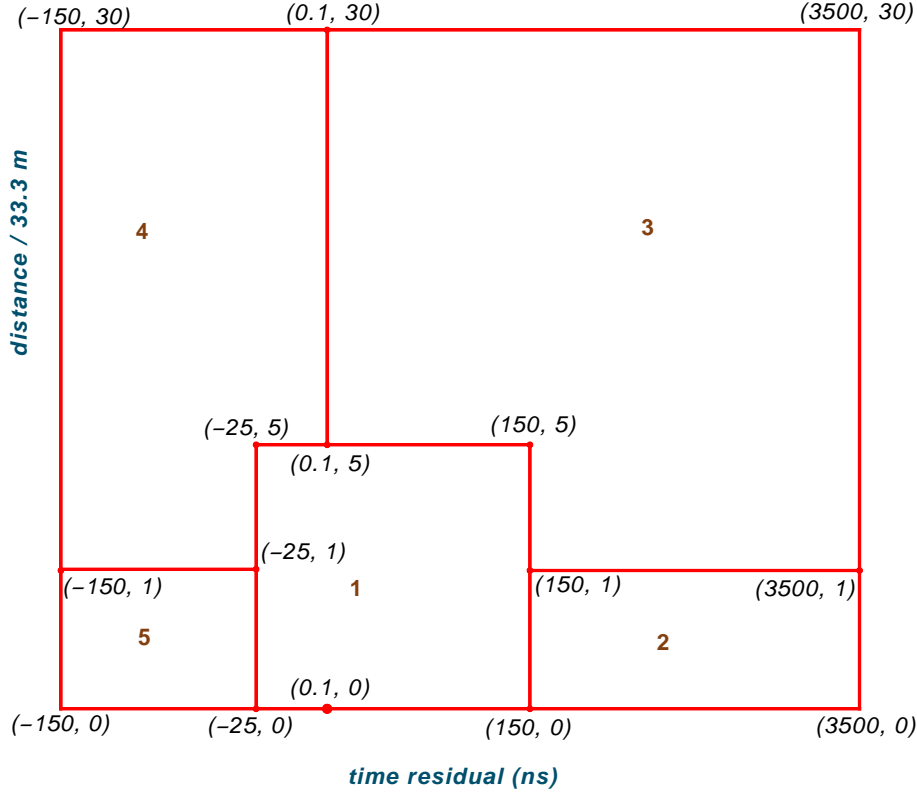}
\end{center}
\caption{Support of $\mathcal{F}_{\sigma}$ for $\sigma=5$~ns.
 Distances and time residuals are restricted to the rectangle $0 \leq \xi \leq 30$
 ($0 \leq d \leq 999$ m) and $-150 \leq t \leq 3500$~ns ($\rho\sigma^{2}=0.1$~ns).}
\label{fig:5nss}
\end{figure}

\begin{figure}[htb] 
\begin{center}
\includegraphics[keepaspectratio,width=12cm]{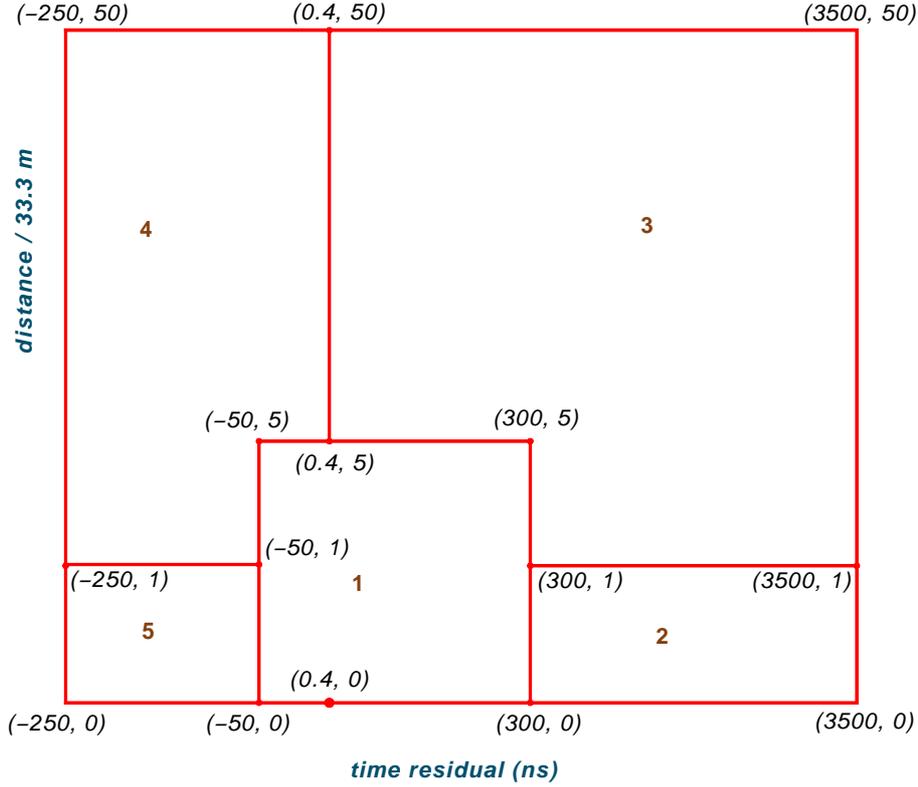}
\end{center}
\caption{Support of $\mathcal{F}_{\sigma}$ for $\sigma=10$~ns.
 Distances and time residuals are restricted to the rectangle $0 \leq \xi \leq 50$
 ($0 \leq d \leq 1665$ m) and $-250 \leq t \leq 3500$~ns ($\rho\sigma^{2}=0.4$~ns).}
\label{fig:10nss}
\end{figure}

Concrete limiting values of the distances and time residuals defining the above five regions depend
on the values of the Pandel PDF parameters and on the value of the time jitter parameter $\sigma$.
Values of $\lambda$ and $\rho$ are given by \eqref{parms} and according to realistic time
jitter values \cite{mureco}, we consider the cases $\sigma=5$ and $\sigma=10$~ns.\\
The corresponding areas of the $\xi-t$ plane where $\mathcal{F}_{\sigma}$
can be evaluated are presented in Figs.~\ref{fig:5nss} and \ref{fig:10nss}.

Outside the large rectangle, numerical evaluation of $\mathcal{F}_{\sigma}$ becomes impossible.
With increasing distance, $\mathcal{F}_{\sigma}$ decreases sharply to values of about
$10^{-200}-10^{-280}$, depending on the value of $t$.
For fixed $\xi$ and increasing $|t|$, computer limitations lead to overflow.\\
To compensate for these effects, one can evaluate $\mathcal{F}_{\sigma}$ on the corresponding
boundary location and impose a penalty value on the resulting log-likelihood value
beyond the regions $1\,-\,5$.

The accuracy while crossing the border of the regions $i$ and $j$ is given by inequality:
\begin{equation}
\label{patch}
r_{ij,\;k}(\xi,\,t)\,\equiv\,\frac{|f_{i}(\xi,\,t)\,-\,f_{j}(\xi,\,t)|}{f_{k}(\xi,\,t)}\,\leq\,10^{-3} ~,
\end{equation}
where $k=i$ or $k=j$.\\
The value of the relative error of about $10^{-3}$ occurs while comparing the function $f_{1}$
with $f_{5}$ and $f_{4}$ with $f_{5}$, i.e. the region with negative $t$ and small $\xi$.
For the remaining areas, the value of $r_{ij}$ is in the range of $10^{-11} \leq r_{ij} \leq 10^{-6}$. 

\section{Derivations of the various approximations for $\mathcal{F}_{\sigma}(\rho,\xi,t)$}

\subsubsection*{Approximation in the region 5: $\xi \leq 1$ and $t \ll -\sigma$}
In this region $\eta \equiv \rho\sigma -t/\sigma$ is positive and we can use the asymptotic
expansion formula {\bf 13.5.1} (bold faced numeration refers to the corresponding sections of \cite{AS})
valid for large $z$ and fixed $a,b$~:
\begin{eqnarray}
\frac{_{1}F_{1}(a,b,z)}{\Gamma(b)} &=&
\frac{e^{i\pi a}z^{-a}}{\Gamma(b-a)}\,
\Biggl[\sum^{R-1}_{n=0}\frac{\Gamma(a+n)\Gamma(1+a-b+n)}{\Gamma(a)\Gamma(1+a-b)n!}(-z)^{-n}
\,+\,\mathcal{O}(z^{-R})\Biggr] \cr\cr\cr
&+&
\frac{e^{z}z^{a-b}}{\Gamma(a)}\,
\Biggl[\sum^{S-1}_{n=0}\frac{\Gamma(b-a+n)\Gamma(1-a+n)}{\Gamma(b-a)\Gamma(1-a)n!}z^{-n}
\,+\,\mathcal{O}(z^{-S})\Biggr] ~.
\label{50}
\end{eqnarray}

Application of \eqref{50} to the combination of hypergeometric functions appearing in
the exact expression \eqref{cpexact} for $\mathcal{F}_{\sigma}(\rho,\xi,t)$~:
\begin{equation}
\frac{_{1}F_{1}(\frac{1}{2}\xi,\frac{1}{2},\frac{1}{2}\eta^{2})}{\Gamma(\frac{1}{2}(\xi+1))}
-\sqrt{2}\,\eta\;
\frac{_{1}F_{1}(\frac{1}{2}(\xi+1),\frac{3}{2},\frac{1}{2}\eta^{2})}{\Gamma(\frac{1}{2}\xi)}
\label{51}
\end{equation}
shows that the rising terms, proportional to $e^{z}$, originating from the second term of \eqref{50},
cancel. For the leading order term we obtain
\begin{equation}
e^{\eta^{2}/2}\,\Biggl[
\frac{\Gamma(\frac{1}{2})\,\eta^{\xi-1}\,2^{(1-\xi)/2}}{\Gamma(\frac{1}{2}\xi)\Gamma(\frac{1}{2}(\xi+1))}
-\frac{\Gamma(\frac{1}{2})\,\eta^{\xi-1}\,2^{(1-\xi)/2}}{\Gamma(\frac{1}{2}\xi)\Gamma(\frac{1}{2}(\xi+1))}
\Biggr]=0
\label{52}
\end{equation}
and the same cancellation occurs for the non-leading order terms.

The remaining terms of \eqref{50}, proportional to $e^{i\pi a}$, give rise (in leading order)
to the expression
\begin{eqnarray}
\frac{e^{i\pi\xi/2}\,\eta^{-\xi}\,2^{\xi/2}}{\Gamma(\frac{1}{2}(\xi+1))\Gamma(\frac{1}{2}(1-\xi))}-
\frac{e^{i\pi(\xi+1)/2}\,\eta^{\xi-1}\,2^{\xi/2}}{\Gamma(\frac{1}{2}\xi)\Gamma(\frac{1}{2}(\xi+1))}=\cr\cr\cr
e^{i\pi\xi/2}\,\eta^{-\xi}\,2^{\xi/2}\,
\Biggl[\frac{1}{\Gamma(\frac{1}{2}(\xi+1))\Gamma(\frac{1}{2}(1-\xi))}-
\frac{e^{i\pi/2}}{\Gamma(\frac{1}{2}\xi)\Gamma(\frac{1}{2}(\xi+1))}\Biggr] ~.
\label{53}
\end{eqnarray}

Using the reflection formulae {\bf 6.1.17}~:
\begin{equation}
\Gamma(z)\Gamma(1-z)=\frac{\pi}{\sin(\pi z)} \text{~~and~~}
\Gamma(z+\frac{1}{2})\Gamma(z-\frac{1}{2})=\frac{\pi}{\cos(\pi z)} ~,
\label{54}
\end{equation}
we can write the leading order expression \eqref{53} as
\begin{eqnarray}
\frac{e^{i\pi\xi/2}\,\eta^{-\xi}\,2^{\xi/2}}{\pi}\,
\biggl(\cos(\pi\xi/2)\,-\,e^{i\pi/2}\sin(\pi\xi/2)\biggr) 
=\frac{\eta^{-\xi}\,2^{\xi/2}}{\pi} ~.
\label{55}
\end{eqnarray}
So, the imaginary part is vanishing, as is expected for a PDF.\\
Using \eqref{55} in the exact expression \eqref{cpexact} for $\mathcal{F}_{\sigma}(\rho,\xi,t)$
results in the leading order approximation \eqref{ll} used in the region 5.

\subsubsection*{Approximation in the region 2: $\xi \leq 1$ and $t \gg \sigma$}
Also here we derive only the leading order approximation using \eqref{50}.\\
When $t > \rho\sigma^{2}$, $\eta$ is negative and for convenience we introduce a positive $\mu$~:
\begin{equation}
\label{57}
\mu \equiv -\eta = \frac{t}{\sigma}-\rho\sigma >0 ~.
\end{equation}
The approximation in the region 2 is obtained by applying \eqref{50} to the exact expression
\eqref{cpexact} for $\mathcal{F}_{\sigma}(\rho,\xi,t)$, written in terms of this positive $\mu$.
The net effect is that now the sign in \eqref{51} between the two $_{1}F_{1}$'s is reflected.
Lengthy but straightforward algebra leads to
\begin{equation}
\label{58}
\mathcal{F}_{\sigma}(\rho,\xi,t) \approx
\frac{\rho^{\xi}\,\sigma^{\xi-1}\,e^{\rho^{2}\sigma^{2}/2-\rho t}\,2^{1-\xi}\,\Gamma(\frac{1}{2})}
{\Gamma(\frac{1}{2}\xi)\Gamma(\frac{1}{2}(1+\xi))}
\biggl(\frac{t-\rho\sigma^{2}}{\sigma}\biggr)^{\xi-1} ~.
\end{equation}
Using the duplication formula {\bf 6.1.18}~:
\begin{equation}
\label{59}
\Gamma(2z)\,=\,2^{2z-1}\pi^{-1/2}\Gamma(z)\Gamma(z+\frac{1}{2})
\end{equation}
we arrive at the approximation in the region 2~:
\begin{equation}
\label{590}
\mathcal{F}_{\sigma}(\rho,\xi,t) \approx
e^{\rho^{2}\sigma^{2}/2}\,\frac{\rho^{\xi}t^{\xi-1}}{\Gamma(\xi)}\,e^{-\rho t} ~.
\end{equation}

In deriving the approximations for the regions 2 and 5 the non-leading terms,
proportional to $(\rho\sigma^{2}/t)^{-n}$, have been neglected.
The corresponding numerical values contributing to the correction factors amount to less than $0.1$\%.
Note that the asymptotic expansion formula \eqref{50} allows for the accounting of the non-leading terms
in a straightforward way if the necessity arises.

\subsubsection*{Approximations in the regions 3 and 4}
To proceed further it is convenient to express the PDF $\mathcal{F}_{\sigma}(\rho,\xi,t)$ in terms of
the parabolic cylinder function $U(a,\,x)$ \cite{AS}.
Comparison with {\bf 13.6.15} and {\bf 13.6.16} shows that the PDF \eqref{cpexact} can be expressed as
\begin{equation}
\label{591}
\mathcal{F}_{\sigma}(\rho,\xi,t)=\frac{(\rho\sigma)^{\xi}}{\sqrt{2\pi\sigma^{2}}}
\,\exp\biggl(\frac{\eta^{2}}{4}-\frac{t^{2}}{2\sigma^{2}}\biggr)
\,U\biggl(\xi-\frac{1}{2},\eta\biggr) ~.
\end{equation}

However, the function $U$ is not implemented in GSL, therefore we can not use the above expression
in the region of small $\xi$ and small $\pm t$.
This disadvantage is compensated by the fact that manipulation with $U$ leads to more compact
expressions than those with the superposition of the two hypergeometric functions.
E.g., it is straightforward to verify that the asymptotic expansion formula for $U$, {\bf 19.8.1},
immediately leads to the results for the regions 2 and 5, and there is no need to follow up the
intermediate steps of calculation in order to see explicitly that the rising terms cancel as we saw
in \eqref{52}, or that the result contains no imaginary part, as we saw in \eqref{55}.\\
Let us consider the region 4, where $t \leq \rho\sigma^{2}$, i.e. where
\begin{equation}
\label{U}
\eta=\rho\sigma-\frac{t}{\sigma} \geq 0 ~.
\end{equation}
Below we list the essential steps necessary to obtain the approximation.\\
Introduction of a new variable $z$,
\begin{equation}
\label{U1}
z \equiv \frac{\eta}{\sqrt{4\xi-2}} ~,
\end{equation}
transforms the equation for a parabolic cylinder function {\bf 19.1.2} into
\begin{equation}
\label{U2}
\frac{{\rm d}^{2}}{{\rm d}z^{2}}
U\biggl(\xi-\frac{1}{2},z\biggr)=(2\xi-1)^{2}\,(1+z^{2})\,U\biggl(\xi-\frac{1}{2},z\biggr) ~.
\end{equation}
Next, we introduce a function $U_{1}$ which is defined as
\begin{equation}
\label{U3}
U_{1} \equiv (1+z^{2})^{1/4}\,U\biggl(\xi-\frac{1}{2},z\biggr)
\end{equation}
and define the variables $\phi$ and $k$ as
\begin{equation}
\label{U3b}
\phi \equiv \sinh^{-1}(z) ~, \quad
k \equiv \frac{1}{2}\biggl[\sinh(\phi)\cosh(\phi)+\ln\biggl(\sinh(\phi)+\cosh(\phi)\biggr)\biggr] ~.
\end{equation}
Finally, we introduce a function $\Psi$ and variable $\beta$ as
\begin{equation}
\label{U4}
\Psi \equiv e^{k(2\xi-1)}\,U_{1} ~, \qquad
\beta \equiv \frac{1}{2}(\tanh(\phi)-1)=\frac{1}{2}\,\biggl(\frac{z}{\sqrt{z^{2}+1}}-1\biggr) ~.
\end{equation}
Making use of the relations
\begin{equation}
\label{U5}
\frac{{\rm d}k}{{\rm d}\phi}=\cosh(\phi)=\sqrt{z^{2}+1} \text{~~and~~}
\frac{{\rm d}k}{{\rm d}\beta}=2\cosh^{4}(\phi)=\frac{1}{8\beta^{2}(1+\beta)^{2}}
\end{equation}
it is straightforward to verify that the function $\Psi$ satisfies the following equation~:
\begin{equation}
\label{U6}
\frac{{\rm d}}{{\rm d}\beta}\,\biggl[\beta^{2}(1+\beta)^{2}\frac{{\rm d}\Psi}{{\rm d}\beta}\,\biggr]-
\frac{(2\xi-1)}{4}\,\frac{{\rm d}\Psi}{{\rm d}\beta}+\frac{20\beta^{2}+20\beta+3}{16}\,\Psi=0 ~.
\end{equation}
Expanding $\Psi$ in inverse powers of $2\xi-1$, we obtain~:
\begin{equation}
\label{U7}
\Psi=1+\frac{a_{1}(\beta)}{2\xi-1}+\frac{a_{2}(\beta)}{(2\xi-1)^{2}}+\cdots
\end{equation}
The first term of the expansion \eqref{U7} amounts to  one, because we have introduced
the transformation $U(\xi-\frac{1}{2},z) \rightarrow U_{1} \rightarrow \Psi$.

The expansion coefficients $a_{i}(\beta)$ of \eqref{U7} are obtained by 
using the expansion \eqref{U7} of $\Psi$ in the equation \eqref{U6}.
Neglecting the terms $\mathcal{O}((2\xi-1)^{-1})$, i.e. retaining only constant terms,
we obtain the following relation for $a_{1}(\beta)$~:
\begin{equation}
\label{U8}
\frac{{\rm d}a_{1}(\beta)}{{\rm d}\beta}=\frac{20\beta^{2}+20\beta+3}{4} ~.
\end{equation}
The corresponding solution is given by
\begin{equation}
\label{U9}
a_{1}(\beta)=\frac{\beta}{12}(20\beta^{2}+30\beta+9) ~,
\end{equation}
which is exactly the numerator $N_{1}$ in the expression for the approximation of the PDF in region 4.
Analyzing in the same fashion the $(2\xi-1)^{-1}$ terms, we obtain the expression for $a_{2}(\beta)$,
which corresponds to $N_{2}$ in the approximation expression of the PDF in region 4.\\
The approximation \eqref{leftdef} of the PDF for the region 4 is obtained by
collecting the various terms together, transforming back to the parabolic cylinder function
$\Psi \rightarrow U_{1} \rightarrow U(\xi-\frac{1}{2},z)$ and using \eqref{U1} and \eqref{591}.

The approximation \eqref{rightdef} in the region 3 is obtained using the same strategy as the one
used for the region 4.
The difference we need to account for is that now $\eta=\rho\sigma-t/\sigma$ is negative which
results in a slightly different series for $\Psi$. 

\section{Summary}
Based on approximations given by eqs.~\eqref{101}-\eqref{ll} we have evaluated the values of
a Gauss convoluted Pandel (CPandel) PDF in an area of the distance-time residual plane which covers
basically all physically relevant parameters to perform track reconstruction in a neutrino telescope.
The approximations are obtained by considering analytical continuations of the exact
expression \eqref{cpexact} for the CPandel PDF in different areas of the $\xi-t$ plane
and, wherever necessary, expanding the result in a numerically convergent series.
As such, our approximations are obtained without involving any new ad hoc
functions or parameters.

The concrete values of $\xi$ and $t$ which define the borders of the regions where each of the
approximations is applicable, are defined by values of the parameters of the original Pandel PDF.
In this report we have used the values given in \eqref{parms}.
Since the expressions \eqref{101}-\eqref{ll} are derived analytically, it is a matter of
straightforward interpolation to define support for a CPandel PDF for different parameter values.

\begin{ack}
The authors would like to thank David Boersma and  Mathieu Ribordy for valuable comments.
This work was supported by the National Science Foundation (NSF-G067771) and  the Netherlands Organisation for Scientific Research (NWO). 
\end{ack}

\end{document}